\documentclass[11pt,a4paper]{article}

\usepackage{amssymb}

\usepackage[dvips]{graphicx}

\begin{document}

\title{\bf The NSVZ $\beta$-function in supersymmetric theories
with different regularizations and renormalization prescriptions}

\author{
A.L.Kataev\\
{\small{\em Institute for Nuclear Research of the Russian Academy of Science,}}\\
{\small {\em 117312, Moscow, Russia}},\\
\\
K.V.Stepanyantz\\
{\small{\em Moscow State University}}, {\small{\em
Faculty of Physics, Department  of Theoretical Physics}}\\
{\small{\em 119991, Moscow, Russia}}}

\maketitle

\begin{abstract}
We briefly review the calculations of quantum corrections related
with the exact NSVZ $\beta$-function in ${\cal N}=1$
supersymmetric theories, paying especial attention to the scheme
dependence of the results. It is explained, how the NSVZ relation
is obtained for the renormalization group functions defined in
terms of the bare coupling constant if a theory is regularized by
higher derivatives. Also we describe, how to construct a special
renormalization prescription which gives the NSVZ relation for the
renormalization group functions defined in terms of the
renormalized coupling constant exactly in all orders for Abelian
supersymmetric theories, regularized by higher derivatives. The
scheme dependence of the NSVZ $\beta$-function (for the
renormalization group functions defined in terms of the
renormalized coupling constant) is discussed in the non-Abelian
case. It is shown that in this case the NSVZ $\beta$-function
leads to a certain scheme-independent equality.
\end{abstract}

\unitlength=1cm

%%%%%%%%%%%%%%%%%%%%%%%%%%%%%%%%%%%

\vspace*{-16.7cm}

\begin{flushright}
{\it Dedicated to the 75-th Birthday\\ of Prof. A.A.Slavnov}
\end{flushright}

\begin{flushright}
INR-TH-2014-012
\end{flushright}

\vspace*{14cm}

\section{Introduction}
\hspace{\parindent}

${\cal N}=1$ supersymmetric gauge theories are interesting from
both theoretical and phenomenological point of view. The exact
NSVZ $\beta$-function relates the $\beta$-function of these
theories with the anomalous dimension of the matter superfields.
For theories containing chiral matter superfields the NSVZ
relation has the following form

\begin{equation}\label{NSVZ}
\beta(\alpha) = - \frac{\alpha^2\Big(3 C_2 - T(R) + C(R)_i{}^j
\gamma_j{}^i(\alpha)/r \Big)}{2\pi(1- C_2\alpha/2\pi)},
\end{equation}

\noindent where the variable $\alpha$ denotes the argument in this
relation. Depending on the definition of the renormalization group
functions, this argument can be either the renormalized coupling
constant, or the bare coupling constant. In Eq. (\ref{NSVZ}) the
following notation is used:

\begin{eqnarray}\label{Notation}
&& \mbox{tr}\,(T^A T^B) \equiv T(R)\,\delta^{AB};\qquad
(T^A)_i{}^k
(T^A)_k{}^j \equiv C(R)_i{}^j;\nonumber\\
&& f^{ACD} f^{BCD} \equiv C_2 \delta^{AB};\qquad\quad r\equiv
\delta_{AA},
\end{eqnarray}

\noindent where $C_2$ and $C(R)_i{}^j$ are the quadratic Casimir
operators, $2 T(R)$ is the Dynkin index of the representation $R$,
and $r$ is a dimension of the gauge group. For the pure ${\cal
N}=1$ Yang--Mills theory the expression (\ref{NSVZ}) gives the
exact $\beta$-function, which appears to be a geometric
progression.

The NSVZ $\beta$-function was constructed in Refs.
\cite{Novikov:1983uc,Novikov:1985rd} using arguments based on the
structure of instanton constributions, namely, by requiring their
renormalization group invariance (for a review, see Ref.
\cite{Shifman:1999mv}). Also it was derived in Refs.
\cite{Jones,Shifman:1986zi} using the arguments based on
anomalies. It is known that in supersymmetric theories the axial
anomaly and the anomaly of the energy-momentum trace belong to the
same supermultiplet. The axial anomaly is contributed only by the
one-loop approximation, while the anomaly of the energy-momentum
trace is proportional to a $\beta$-function. Accurately analyzing
these facts, it is possible to obtain the NSVZ $\beta$-function.
This analysis is rather nontrivial. For example, according to Ref.
\cite{ArkaniHamed:1997mj} the multiplication of the matter
superfields by the renormalization constant $Z$ changes the
divergences in supersymmetric theories. This rescaling anomaly can
be used for explaining the origin of the higher order corrections
in the NSVZ $\beta$-function. In the Abelian case the NSVZ
expression was first derived in Refs.
\cite{Vainshtein:1986ja,Shifman:1985fi}. In Ref.
\cite{Kraus:2002nu} the NSVZ $\beta$-function was found using the
non-renormalization theorem for the topological term and the
Slavnov--Taylor identities \cite{Taylor:1971ff,Slavnov:1972fg}.

The NSVZ $\beta$-function relates different divergent
contributions to the effective action in an interesting and
nontrivial way. In particular, it can be used for proving the
finiteness of ${\cal N}=2$ supersymmetric theories beyond the
one-loop approximation
\cite{Novikov:1983uc,Shifman:1999mv,Buchbinder:2014wra}. The NSVZ
$\beta$-function can be also used for investigating the
possibility of constructing finite ${\cal N}=1$ supersymmetric
gauge theories \cite{Shifman:1999mv}. The exact relations similar
to the NSVZ $\beta$-function can be written in theories with the
soft supersymmetry breaking for the renormalization of the gaugino
mass \cite{Hisano:1997ua,Jack:1997pa,Avdeev:1997vx}. However, in
this paper we will not discuss them.

Although the NSVZ $\beta$-function was constructed a long time
ago, it was not clear, how it can be derived using the explicit
calculations of Feynman diagrams and what renormalization
prescription should be used in order to obtain it. Thus, up to now
investigation of the NSVZ relation remains an actual problem. In
this paper we briefly review various calculations which are
related with this problem and describe the scheme in which the
NSVZ $\beta$-function is obtained with the higher derivative
regularization, proposed by A.A.Slavnov
\cite{Slavnov:1971aw,Slavnov:1972sq}, in the Abelian case.
Moreover, we find how the NSVZ $\beta$-function is changed under
the finite renormalizations in the non-Abelian case and derive a
scheme independent consequence of the NSVZ relation.

\section{The NSVZ relation for the renormalization group functions
defined in terms of the bare coupling constant}
\hspace{\parindent}

One of the first derivations of the NSVZ $\beta$-function was made
in Ref. \cite{Novikov:1983uc} by requiring the renormalization
group invariance of the instanton contributions. It should be
noted that in Ref. \cite{Novikov:1983uc} the NSVZ relation was
written for the $\beta$-function defined in terms of the bare
coupling constant. The same definition was also used in Refs.
\cite{Novikov:1985rd,Vainshtein:1986ja,Shifman:1985fi,ArkaniHamed:1997mj,Shifman:1985vu}.

In terms of the bare coupling constant the $\beta$-function and
the anomalous dimension are defined as

\begin{eqnarray}\label{Beta_Bare}
&& \beta\Big(\alpha_0(\alpha,\Lambda/\mu)\Big) \equiv \frac{d
\alpha_0(\alpha,\Lambda/\mu)}{d\ln\Lambda}
\Big|_{\alpha=\mbox{\scriptsize const}};\vphantom{\Bigg|}\\
\label{Gamma_Bare} &&
\gamma_i{}^j\Big(\alpha_0(\alpha,\Lambda/\mu)\Big) \equiv -
\frac{d \ln Z_i{}^j(\alpha,\Lambda/\mu)}{d
\ln\Lambda}\Big|_{\alpha=\mbox{\scriptsize const}},
\end{eqnarray}

\noindent where $\Lambda$ is a dimensionful regularization
parameter and $\mu$ is a renormalization scale. For calculating
these renormalization group functions the left hand side is
considered as a function of the bare coupling constant and the
limit $\Lambda\to \infty$ is not taken. In addition to the
regularization, it is necessary to choose a prescription for
constructing the function $\alpha(\alpha_0,\Lambda/\mu)$ and the
renormalization constant $Z_i{}^j$. By other words, it is
necessary to choose a renormalization scheme.

The renormalization group functions (\ref{Beta_Bare}) and
(\ref{Gamma_Bare}) considered as functions of the bare coupling
constant are scheme independent, because it is possible to express
them in terms of the unrenormalized Green functions. Nevertheless,
these functions depend on the regularization. However, in the
early papers this dependence was not investigated. A different
definition of the renormalization group functions (in terms of the
renormalized coupling constant) was used in Refs.
\cite{Shifman:1986zi,Kraus:2002nu}.

The NSVZ relation for the renormalization group functions
(\ref{Beta_Bare}) and (\ref{Gamma_Bare}) appeared again in papers
devoted to calculations of quantum corrections with the higher
covariant derivative regularization
\cite{Slavnov:1971aw,Slavnov:1972sq}. This regularization also
includes insertion of Pauli--Villars determinants for removing the
one-loop divergences \cite{Faddeev:1980be}. It is consistent and
(in supersymmetric theories) can be formulated in a manifestly
supersymmetric way \cite{Krivoshchekov:1978xg,West:1985jx}.
However, it produces complicated loop integrals. Due to this
reason explicit calculation of the one-loop $\beta$-function for
the (non-supersymmetric) Yang--Mills theory with the higher
covariant derivative regularization \cite{Martin:1994cg} was made
much later than the calculation with the dimensional
regularization \cite{Gross:1973id,Politzer:1973fx}. After
corrections made in \cite{Asorey:1995tq,Bakeyev:1996is} the
results for the $\beta$-function in both regularizations coincide,
as it should be \cite{Pronin:1997eb}.\footnote{The result of Ref.
\cite{Martin:1994cg} was different, the correct result for the
one-loop $\beta$-function was found later.} In supersymmetric
theories regularized by higher covariant derivatives the integrals
defining the $\beta$-function (\ref{Beta_Bare}) appear to be
integrals of total derivatives
\cite{Soloshenko:2003nc,Pimenov:2009hv} and even integrals of
double total derivatives \cite{Smilga:2004zr,Stepanyantz:2011bz}.
This allows to calculate one of loop integrals analytically and to
obtain the NSVZ relation between the $\beta$-function and the
anomalous dimensions of the matter superfields. For the Abelian
supersymmetric gauge theories this feature was proved exactly in
all loops in Ref. \cite{Stepanyantz:2011jy} by explicit summation
of supergraphs and in Ref. \cite{Stepanyantz:2014ima} using a
method based on the Schwinger--Dyson equations. Calculations with
the higher covariant derivative regularization for the non-Abelian
supersymmetric gauge theories were made only in the one- and
two-loop approximations \cite{Pimenov:2009hv}. They reveal exactly
the same features as in the Abelian case, namely, the
factorization of integrals for the $\beta$-function
(\ref{Beta_Bare}) into integrals of (double) total derivatives.
This factorization is a reason for appearing the NSVZ relation in
supersymmetric theories for the renormalization group functions
defined in terms of the bare coupling constant. This feature takes
place in case of using the higher covariant derivative
regularization, and it is not so far clear, if it is valid for
other regularizations. Thus, the choice of the higher covariant
derivative regularization is very important.

\section{Scheme dependence of the NSVZ $\beta$-function}
\hspace{\parindent}

In quantum field theory the renormalization group functions are
usually defined in a different way \cite{Bogolyubov:1980nc}, in
terms of the renormalized coupling constant:

\begin{eqnarray}
\label{Beta_Renormalized} &&
\widetilde\beta\Big(\alpha(\alpha_0,\Lambda/\mu)\Big) \equiv
\frac{d\alpha(\alpha_0,\Lambda/\mu)}{d\ln\mu}\Big|_{\alpha_0=\mbox{\scriptsize const}};\\
\label{Gamma_Renormalized} &&
\widetilde\gamma_i{}^j\Big(\alpha(\alpha_0,\Lambda/\mu)\Big)
\equiv \frac{d\ln Z_i{}^j(\alpha(\alpha_0,\Lambda/\mu),
\Lambda/\mu)}{d\ln\mu}\Big|_{\alpha_0=\mbox{\scriptsize
const}}.\qquad
\end{eqnarray}

\noindent It is well known that the renormalization group
functions (\ref{Beta_Renormalized}) and (\ref{Gamma_Renormalized})
are scheme dependent unlike the functions (\ref{Beta_Bare}) and
(\ref{Gamma_Bare}).

There are a lot of papers in which the renormalization group
functions (\ref{Beta_Renormalized}) and (\ref{Gamma_Renormalized})
are calculated for supersymmetric theories in the low orders of
the perturbation theory. The exact NSVZ $\beta$-function can be
compared with the results of these calculations. Usually, most of
perturbative calculations in non-supersymmetric theories are made
using the dimensional regularization
\cite{'tHooft:1972fi,Bollini:1972ui,Ashmore:1972uj,Cicuta:1972jf}.
This regularization was also used for calculating the
$\beta$-function in supersymmetric theories in one-
\cite{Ferrara:1974pu} and two-loop \cite{Jones:1974pg}
approximations. (In these approximations the $\beta$-function is
scheme-independent, so that the result is the same as in the case
of using other regularizations, in particular, the dimensional
reduction \cite{Siegel:1979wq}.) Since the dimensional
regularization breaks the supersymmetry, it is not convenient to
use it in supersymmetric theories.\footnote{In general, using of
non-invariant regularizations is possible, if they are
supplemented by special renormalization prescriptions which
restore the Slavnov--Taylor identities
\cite{Slavnov:2001pu,Slavnov:2002ir,Slavnov:2002kg,Slavnov:2003cx}.}
That is why supersymmetric theories are mostly regularized by the
dimensional reduction, which is not, however, mathematically
consistent \cite{Siegel:1980qs}. Removing of the inconsistencies
cannot be made without loss of supersymmetry
\cite{Avdeev:1981vf,Avdeev:1982xy}. Nevertheless, the calculations
with the dimensional reduction in the
$\overline{\mbox{DR}}$-scheme (which is a modification of the
$\overline{\mbox{MS}}$-scheme \cite{Bardeen:1978yd}) were made in
various supersymmetric theories. In the ${\cal N}=4$
supersymmetric Yang--Mills theory the three- \cite{Avdeev:1981ew}
and four-loop \cite{Velizhanin:2010vw} calculations demonstrate
vanishing of the $\beta$-function. Moreover, the dimensional
reduction does not break supersymmetry even in the four-loop
approximation \cite{Velizhanin:2010vw}, although it is possible in
higher loops. The $\beta$-function of the pure ${\cal N}=2$
supersymmetric Yang--Mills theory vanishes at the three-loops
\cite{Avdeev:1981ew}, but supersymmetry is broken in this
approximation \cite{Avdeev:1982np,Velizhanin:2008rw}. (The
calculation in Ref. \cite{Avdeev:1982np} was corrected in
\cite{Velizhanin:2008rw}.) At present, it is still not clear
whether this is related with the inconsistency of the dimensional
reduction, or with using a non-supersymmetric gauge condition, or
with some other reason. For ${\cal N}=1$ supersymmetric theories a
$\beta$-function was calculated in the three-
\cite{Avdeev:1981ew,Jack:1996vg} and four-loop
\cite{Harlander:2006xq,Jack:2007ni} approximations, see Ref.
\cite{Mihaila:2013wma} for a recent review. The results do not
agree with the NSVZ $\beta$-function starting from the three-loop
approximation, where the scheme dependence becomes essential. Some
consequences of this difference are discussed in Ref.
\cite{Ryttov:2012qu}. However, the NSVZ $\beta$-function can be
obtained after a special tuning of the subtraction scheme which
should be made in every order of the perturbation theory
\cite{Jack:1996vg,Jack:1996cn,Jack:1998uj}. The corresponding
finite renormalization of the coupling constant can be found by
investigating theories with the softly broken supersymmetry
\cite{Jack:1998uj}. Therefore, there is a question, how to
formulate a prescription which gives the NSVZ scheme in
supersymmetric theories in all orders of the perturbation theory.
The answer can be found using a different regularization for which
the NSVZ scheme can be formulated explicitly. Note that other
regularizations were used mainly for calculations in the one- and
two-loop approximations. We can mention Ref. \cite{Mas:2002xh} in
which the two-loop $\beta$-function of the ${\cal N}=1$
supersymmetric Yang--Mills theory was calculated using a version
of the differential renormalization \cite{Freedman:1991tk},
proposed in \cite{delAguila:1998nd}, which does not break
supersymmetry \cite{delAguila:1997yd}. In Ref.
\cite{Shifman:1985tj} the two-loop $\beta$-function for the ${\cal
N}=1$ supersymmetric electrodynamics was found by using the
operator product expansion and a cutoff in the coordinate space.
Since a two-loop $\beta$-function is scheme independent in
theories with a single coupling constant, these results cannot be
used for investigating the scheme dependence. For completeness, we
also mention a regularization based on using ${\cal N}=4$ or
finite ${\cal N}=2$ supersymmetric Yang--Mills theories and
introducing a cutoff breaking supersymmetry to ${\cal N}=1$. This
regularization was used in Ref. \cite{ArkaniHamed:1997mj}.
However, it cannot be applied for general ${\cal N}=1$
supersymmetric theories.

Quantum corrections also appear in instanton calculations
\cite{'tHooft:1976fv}. Finding them one again encounters a scheme
dependence of the results. In supersymmetric theories instanton
contributions were obtained taking into account quantum
corrections up to the two-loop order \cite{Morris:1985pb} using
the regularization by the dimensional reduction. The result
confirmed the argumentation proposed in Ref. \cite{Novikov:1983uc}
for derivation of the NSVZ $\beta$-function, namely, vanishing of
quantum corrections to the vacuum energy in the instanton
background. This implies that the calculation of quantum
corrections near the instanton background agrees with the NSVZ
expression at the two-loop level. Although in this approximation
the $\beta$-function defined in terms of the renormalized coupling
constant is scheme-independent, the investigation of the instanton
correction structure reveals a possibility of the
scheme-dependence in higher orders
\cite{Morris:1985pt,Morris:1986su}. Therefore, a similar problem
of the scheme-dependence is also present for instanton
calculations. Unfortunately, explicit instanton calculations which
take into account three-loop corrections are not yet made.

\section{NSVZ scheme}
\hspace{\parindent}

In all orders the NSVZ scheme has been naturally constructed using
the higher covariant derivative regularization at least in the
Abelian case \cite{Kataev:2013eta,Kataev:2013csa}. As we have
already mentioned above, with the dimensional reduction there is
no general prescription which allows to obtain such a scheme in
all orders. However, if the ${\cal N}=1$ supersymmetric
electrodynamics with an arbitrary ($N_f$) number of flavors is
regularized by higher derivatives, this can be done by a special
choice of the subtraction scheme. Namely, the NSVZ scheme is
constructed by imposing the boundary conditions

\begin{equation}
Z_3(\alpha,x_0)=1;\qquad Z(\alpha,x_0)=1
\end{equation}

\begin{table}
\begin{center}
\begin{tabular}{|c|c|c|c|c|}
\hline Subtraction scheme$\vphantom{\Big(}$ & 1-loop & 2-loop &
\multicolumn{2}{|c|}{3-loop} \\
\hline Overall factor$\vphantom{\Big(}$ & $\alpha^2 N_f/\pi$ &
$\alpha^3 N_f/\pi^2$
& $\alpha^4 N_f/\pi^3$ & $\alpha^4 (N_f)^2/\pi^3$ \\
\hline $\overline{\mbox{DR}}$ $\vphantom{\Big(}$ & $1$
& $1$ & $-1/2$ & $-3/4$ \\
\hline HD + NSVZ$\vphantom{\Big(}$ & $1$ &
$1$ & $-1/2$ & $- 1-\sum_{I=1}^n c_I \ln a_I$ \\
\hline MOM$\vphantom{\Big(}$ & $1$ &
$1$ & $-1/2$ & $-3(1-\zeta(3))/2$ \\
\hline
\end{tabular}
\end{center}
\caption{$\beta$-function for the ${\cal N}=1$ supersymmetric
electrodynamics with $N_f$ flavors in different subtraction
schemes. (For the MOM scheme the results obtained with the
dimensional reduction and with the higher derivative
regularization coincide.)}\label{Table_Beta}
\end{table}

\begin{table}
\begin{center}
\begin{tabular}{|c|c|c|c|}
\hline Subtraction scheme$\vphantom{\Big(}$ & 1-loop & \multicolumn{2}{|c|}{2-loop} \\
\hline Overall factor$\vphantom{\Big(}$ & $\alpha/\pi$ &
$\ \qquad\alpha^2/\pi^2\qquad\ $ & $\alpha^2 N_f/\pi^2$  \\
\hline $\overline{\mbox{DR}}$ $\vphantom{\Big(}$ & $-1$
& $1/2$ & $1/2$ \\
\hline HD + NSVZ$\vphantom{\Big(}$ & $-1$ &
$1/2$ & $1+\sum_{I=1}^n c_I \ln a_I$ \\
\hline MOM$\vphantom{\Big(}$ & $-1$ &
$1/2$ & $1/2$ \\
\hline
\end{tabular}
\end{center}
\caption{The anomalous dimension of the matter superfields for the
${\cal N}=1$ supersymmetric electrodynamics with $N_f$ flavors in
different subtraction schemes. (For the MOM scheme the results
obtained with the dimensional reduction and with the higher
derivative regularization coincide.)}\label{Table_Gamma}
\end{table}

\noindent on the renormalization constants for the coupling
constant and the matter superfield, respectively, where $x_0$ is a
certain (arbitrary) value of $x=\ln\Lambda/\mu$. In this case the
renormalization group functions (\ref{Beta_Renormalized}) and
(\ref{Gamma_Renormalized}) defined in terms of the renormalized
coupling constant coincide with the renormalization group
functions (\ref{Beta_Bare}) and (\ref{Gamma_Bare}) defined in
terms of the bare coupling constant \cite{Kataev:2013eta}. The
explicit three-loop calculations in this case were made for ${\cal
N}=1$ supersymmetric electrodynamics with $N_f$ flavors
\cite{Kataev:2013csa}. It is expedient to compare the result with
the corresponding calculations made with other regularizations and
subtraction schemes. The results for the $\beta$-function and the
anomalous dimension of the matter superfields are presented in
Tables \ref{Table_Beta} and \ref{Table_Gamma}, respectively. For
the NSVZ scheme obtained with the higher derivative regularization
$c_I$ denote degrees of the Pauli--Villars determinants, which are
introduced in order to cancel one-loop divergences according to
Ref. \cite{Slavnov:1977zf}, and $a_I = M_I/\Lambda$ are ratios of
the Pauli--Villars masses to the parameter in higher derivative
term. (It is essential that $a_I$ should not depend on the bare
coupling constant.) These tables were constructed using the
results of \cite{Jack:1996vg,Kataev:2013csa}. From them we see
that the NSVZ relation for the considered theory,

\begin{equation}
\beta(\alpha) = \frac{\alpha^2
N_f}{\pi}\Big(1-\gamma(\alpha)\Big),
\end{equation}

\noindent is valid in all schemes in one- and two-loop
approximations. Moreover, in the three-loop approximation the
terms proportional to $\alpha^4 N_f$ in the $\beta$-function and
proportional to $\alpha^2 (N_f)^0$ in the anomalous dimension are
the same in all schemes and satisfy the NSVZ relation. This is a
consequence of the scheme-independence of such terms, which was
proved in \cite{Kataev:2013csa} in all orders and is discussed
below. The terms proportional to $\alpha^4 (N_f)^2$ in the
$\beta$-function and the corresponding terms in the anomalous
dimension satisfy the NSVZ relation only in the NSVZ scheme.

\section{Scheme dependence of the NSVZ $\beta$-function in
the Abelian and non-Abelian cases}
\hspace{\parindent}

Terms of the first degree in $N_f$ in the $\beta$-function and
terms without $N_f$ in the anomalous dimension are
scheme-independent not only in ${\cal N}=1$ supersymmetric
electrodynamics with $N_f$ flavors. The same statement is also
valid in the scalar and spinor electrodynamics with $N_f$ flavors.
Really, under the finite renormalizations

\begin{equation}
\alpha \to \alpha'(\alpha);\qquad Z'(\alpha',\Lambda/\mu) =
z(\alpha) Z(\alpha,\Lambda/\mu),
\end{equation}

\noindent the renormalization group functions defined in terms of
the renormalized coupling constant are changed as follows:

\begin{eqnarray}\label{Finite_Renormalization_Beta}
&& \widetilde \beta'(\alpha') =
\frac{d\alpha'}{d\ln\mu}\Big|_{\alpha_0=\mbox{\scriptsize const}}
= \frac{d\alpha'}{d\alpha} \widetilde \beta(\alpha);\\
\label{Finite_Renormalization_Gamma} && \widetilde
\gamma'(\alpha') = \frac{d\ln
Z'{}_i{}^j}{d\ln\mu}\Big|_{\alpha_0=\mbox{\scriptsize const}} =
\frac{d\ln z}{d\alpha}\cdot \widetilde \beta(\alpha) +
\widetilde\gamma(\alpha),
\end{eqnarray}

\noindent and in the case of the ${\cal N}=1$ supersymmetric
electrodynamics with $N_f$ flavors $\widetilde\beta(\alpha)$ and
$\alpha'(\alpha)-\alpha$ are proportional at least to $N_f$. The
scheme-independence of the considered terms is in agreement with
the four- \cite{Gorishnii:1990kd} and five-loop
\cite{Baikov:2012zm} calculations in the $\overline{\mbox{MS}}$
and $\mbox{MOM}$ schemes for the quantum electrodynamics with
$N_f$ flavors.\footnote{For $N_f=1$ the five-loop results of Ref.
\cite{Baikov:2012zm} coincide with the results of Ref.
\cite{Kataev:2012rf}, which confirms their correctness.} Also the
scheme independence of the considered terms can be seen comparing
the three-loop results for the scalar electrodynamics in the
$\overline{\mbox{MS}}$-scheme
\cite{Broadhurst:1995dq,Chetyrkin:1983qc} and the on-shell scheme
\cite{Broadhurst:1995dq}.

For a general non-Abelian supersymmetric Yang--Mills theory finite
renormalizations can also include Yukawa couplings $\lambda^{ijk}$
(which are totally symmetric in indexes $ijk$):

\begin{equation}\label{NonAbelian_Finite_Renormalization}
\alpha \to \alpha'(\alpha,\lambda);\qquad \lambda \to
\lambda'(\alpha,\lambda);\qquad
Z'{}_i{}^j(\alpha',\lambda',\Lambda/\mu) = z_i{}^k(\alpha,\lambda)
Z_k{}^j(\alpha,\lambda,\Lambda/\mu),
\end{equation}

\noindent where we assume that $z$ and $Z$ commute. Then the
generalization of Eqs. (\ref{Finite_Renormalization_Beta}) and
(\ref{Finite_Renormalization_Gamma}) has the form

\begin{eqnarray}\label{Finite_Renormalization_Beta_Non_Abelian}
&& \widetilde \beta'(\alpha',\lambda') =
\frac{\partial\alpha'}{\partial\alpha} \widetilde
\beta(\alpha,\lambda) +\frac{3}{2} \Big(\lambda^{ljk}\,
\widetilde\gamma_l{}^i(\alpha,\lambda)\,
\frac{\partial\alpha'}{\partial \lambda^{ijk}} + \lambda^*_{ljk}\,
\widetilde\gamma_i{}^l(\alpha,\lambda)\,
\frac{\partial\alpha'}{\partial \lambda^*_{ijk}} \Big);\\
\label{Finite_Renormalization_Gamma_Non_Abelian} && \widetilde
\gamma'{}_i{}^j(\alpha',\lambda') =
\widetilde\gamma_i{}^j(\alpha,\lambda) + \frac{\partial\ln
z_i{}^j}{\partial\alpha}\cdot \widetilde
\beta(\alpha,\lambda)\nonumber\\
&& \qquad\qquad\qquad + \frac{3}{2}
\Big(\lambda^{lmn}\,\widetilde\gamma_l{}^k(\alpha,\lambda)\,
\frac{\partial\ln z_i{}^j}{\partial \lambda^{kmn}} +
\lambda^*_{lmn}\, \widetilde\gamma_k{}^l(\alpha,\lambda)\,
\frac{\partial\ln z_i{}^j}{\partial \lambda^*_{kmn}}\Big).
\end{eqnarray}

\noindent Note that deriving this equation we took into account
that according to the non-renormalization theorem
\cite{Grisaru:1979wc} there are no divergent quantum corrections
to the superpotential. As a consequence, renormalization constants
for the Yakawa couplings can be expressed via the renormalization
constants $Z_i{}^j$ for the chiral superfields:

\begin{equation}
\lambda^{ijk} = \lambda_0^{mnp} (Z^{1/2})_m{}^i (Z^{1/2})_n{}^j
(Z^{1/2})_p{}^k,
\end{equation}

\noindent where $\lambda_0$ and $\lambda$ denote the bare and
renormalized Yukawa couplings, respectively. $\lambda'$ is given
by a similar expression which involves the renormalization
constant $Z'$.

Let us assume that in the non-Abelian case the $\beta$-function
$\widetilde\beta(\alpha,\lambda)$ and the anomalous dimension
$\widetilde\gamma_i{}^j(\alpha,\lambda)$ defined in terms of the
renormalized coupling constant satisfy the NSVZ relation. Then we
obtain that after the finite renormalization
(\ref{NonAbelian_Finite_Renormalization})

\begin{eqnarray}\label{New_Beta}
&&\hspace*{-4mm} \widetilde\beta'(\alpha',\lambda') = -
\frac{\alpha^2}{2\pi(1-
C_2\alpha/2\pi)\partial\alpha/\partial\alpha' - \alpha^2
C(R)_l{}^k \partial\ln
z_k{}^l/\partial\ln\alpha'\vphantom{\Big)}}\Bigg\{3 C_2 -
T(R)\nonumber\\
&&\hspace*{-4mm} + \frac{1}{r} C(R)_m{}^n
\Bigg[\widetilde\gamma'{}_n{}^m(\alpha',\lambda') - \frac{3}{2}
\Bigg((\lambda')^{ljk}\,\widetilde\gamma'{}_l{}^i(\alpha',\lambda')\,
\frac{\partial\ln z_n{}^m}{\partial (\lambda')^{ijk}} +
(\lambda')^*_{ljk}\, \widetilde\gamma'{}_i{}^l(\alpha',\lambda')\,
\frac{\partial\ln z_n{}^m}{\partial (\lambda')^*_{ijk}}\Bigg)
\Bigg]\qquad
\nonumber\\
&&\hspace*{-4mm} + \frac{3}{2}\cdot
\frac{2\pi}{\alpha^2}\Big(1-C_2 \frac{\alpha}{2\pi}\Big)
\Bigg((\lambda')^{ljk}\,
\widetilde\gamma'{}_l{}^i(\alpha',\lambda')\,
\frac{\partial\alpha}{\partial (\lambda')^{ijk}} +
(\lambda')^*_{ljk}\, \widetilde\gamma'{}_i{}^l(\alpha',\lambda')\,
\frac{\partial\alpha}{\partial (\lambda')^*_{ijk}} \Bigg)
\Bigg\}_{\alpha=\alpha(\alpha',\lambda')}.
\end{eqnarray}

\noindent This equality can be used for analyzing the scheme
(in)dependence of the results. In particular, we observe that in
$L$ loops the terms proportional to $\mbox{tr}\left(C(R)^L\right)$
are the same in both sides of Eq. (\ref{NSVZ}) for an arbitrary
renormalization prescription. Really, it is evident from Eq.
(\ref{New_Beta}) that modifications of the NSVZ relation for all
terms which do not contain Yukawa couplings can come only from the
denominator. However, all such corrections are proportional either
to $C_2$, $T(R)$, $\lambda$, or to products of at least two
traces. Therefore, the term $\mbox{tr}\left(C(R)^L\right)$, which
contains a single trace cannot be modified. We have verified that
in the three-loop approximation Eq. (\ref{New_Beta}) is in
agreement with the results of Ref. \cite{Jack:1996vg}. In
particular, according to Ref. \cite{Jack:1996vg} in the three-loop
approximation the terms proportional to
$\mbox{tr}\left(C(R)^3\right)$ are the same in the
$\overline{\mbox{DR}}$ and $\mbox{NSVZ}$ schemes.

The results for the ${\cal N}=1$ supersymmetric electrodynamics
with $N_f$ flavors are obtained if we set

\begin{equation}
C_2 = 0;\qquad C(R)_i{}^j = \delta_i{}^j;\qquad T(R) = 2 N_f
\qquad r=1,
\end{equation}

\noindent where the indexes $i$ and $j$, numerating chiral
superfields, take values from 1 to $2N_f$. As a consequence,

\begin{equation}
\mbox{tr}\left(C(R)^L\right) = 2N_f.
\end{equation}

\noindent Therefore, the scheme-independence of the terms
proportional to $\mbox{tr}\left(C(R)^L\right)$ in the Abelian case
leads to the scheme-independence of terms proportional to the
first degree $N_f$, which was proved in Ref.
\cite{Kataev:2013csa}.

\section{Conclusion}
\hspace{\parindent}

Application of the higher covariant derivative regularization to
calculations of quantum corrections in supersymmetric theories
allows to solve the long standing problem of obtaining the NSVZ
$\beta$-function by the direct summation of the supergraphs, at
least, in the Abelian case. The NSVZ relation appears for the
renormalization group functions defined in terms of the bare
coupling constant. If the renormalization group functions are
defined in terms of the renormalized coupling constant,
application of the higher derivative regularization allows to
naturally construct the NSVZ scheme in the Abelian case. Although
quantum corrections in non-Abelian theories have a similar
structure, so far such a scheme is not constructed in this case.
However, we hope that this can be made similarly to the Abelian
theory. In this paper we only investigated the scheme dependence
of the NVSZ relation for the renormalization group functions
defined in terms of the renormalized coupling constant in the
non-Abelian case. It is shown that the terms proportional to
$\mbox{tr}\left(C(R)^L\right)$, where $L$ is a number of loops,
appeared to be scheme-independent. As a consequence, such terms in
the left and right sides of Eq. (\ref{NSVZ}) coincide for an
arbitrary renormalization prescription.

\bigskip
\bigskip

\noindent {\Large\bf Acknowledgements.}

\bigskip

\noindent  The  work of A.L.Kataev is   supported  in part by the
Grant NSh-2835.2114.2 and Russian Foundation of Basic Research
grant No 14-01-00647. The work of K.V.Stepanyantz was supported by
Russian Foundation for Basic Research grant No 14-01-00695.

\end{document}